\documentclass[aps,prl,reprint]{revtex4-2}

\usepackage{graphicx}
\usepackage{dcolumn}
\usepackage{bm}
\usepackage{amsmath}
\usepackage{xcolor}
\usepackage{ulem}

\begin{document}

\preprint{APS/123-QED}

\title{Hydrodynamic thinning of a coating film induced by a small solid defect: evidence of a time-minimum thickness}
\author{Alice Etienne-Simonetti}
 \author{Frédéric Restagno}
\author{Emmanuelle Rio}%
 \email{emmanuelle.rio@universite-paris-saclay.fr}
\affiliation{%
Université Paris-Saclay, CNRS, Laboratoire de Physique des Solides, 91405, Orsay, France.
}%
\author{Isabelle Cantat}
\affiliation{Institute of Physics of Rennes, Rennes, France}%

\date{\today}

\begin{abstract}

During coating processes, dust deposition can lead to an uneven thickness in the resulting film, posing significant problems in industrial processes. 
Our study explores the effects of solid defects using a vertical cylindrical fiber deposited on a silicone oil film coating a horizontal solid substrate. 
We use a hyperspectral camera to measure the film thickness by interferometry in the vicinity of the defect. 
As predicted and observed in many studies in various geometries, a circular groove appears around the fiber because of the capillary suction induced by the meniscus that grows at the bottom of the fiber. 
We measure the evolution of the thickness of the film at the groove over time. The thickness decreases before increasing again leading to the healing of the perturbation at long time.
We propose that healing is due to the arrest of the suction when the meniscus reaches its equilibrium shape.
By combining geometric analysis with the thin film equation, we have developed scaling laws that predict both the minimum thickness of the groove, that we call the time-minimum thickness, and the time required to reach this minimum. 
If the time-minimum thickness reaches the thickness at which intermolecular forces begin to play a role prior to healing, the thickness of the groove will stop decreasing and saturate due to the competition between drainage and repulsive intermolecular forces.
Based on the previous scaling law, we developed a scaling law accounting for the critical initial thickness of the film below which the intermolecular repulsion will start to have an effect, which is in good agreement with our experiments.
These results thus offer valuable insights into predicting and preventing defects in coating processes, thereby improving the quality and reliability of coated products in various industries.

\end{abstract}

\maketitle

\section{Introduction} 
Surface coating refers to the process of applying a layer or layers of a substance onto a surface to protect, improve, or decorate it. 
This process is used to enhance properties like the aspect of the surface, its corrosion, wear or scratch resistance, and its chemical stability.  
Among different possible coating techniques, such as chemical vapor deposition (CVD), physical vapor deposition (PVD), and electroplating, the use of a liquid is a cheap and common technique \cite{ruschak1985coating}. 
In this technique, surface coating typically involves spreading a liquid film, followed by a drying process \cite{oron1997long,schweizer2012liquid}.
Most applications, designed to confer chemical, optical, as well as mechanical properties to the surface necessitate a film that is homogeneous in thickness at an optical scale.
Nevertheless, spontaneous instabilities, such as the ribbing instability can alter the film homogeneity by creating ribs while roll-coating \cite{ruschak1985coating,lopez2002non} and various mechanisms can induce thickness heterogeneities through drying such as evaporation-induced instabilities \cite{magdelaine2019hydrodynamique,robert2022hydrodynamique}, evaporation inhomogeneities \cite{faidherbe2023drying} or the coffee ring effect \cite{deegan1997capillary,kim2016controlled,corpart2023coffee} in which particles set on the edges of the drop.

The presence of a defect at the liquid/air interface can increase these heterogeneities \cite{jalaal2019capillary,hack2018printing,garcia2023drawing}.
When a wetting sub-millimetric defect, such as a dust particle, is deposited on a film of micrometric thickness $h_0$, the meniscus rises around the dust particle.
This phenomenon can be reproduced using a fiber of small radius $r_{\rm{f}}$ deposited vertically in contact with a liquid film coating a horizontal solid substrate.
The equilibrium profile, which corresponds to the long-time situation, is a flat film connecting a meniscus.
When the fiber radius is small compared to the capillary length, the equilibrium meniscus height scales as the fiber radius \cite{quere1994meniscus,james1974meniscus} and the thickness profile decreases continuously from this maximum to the flat film thickness $h_0$.

Upon contact between the fiber and the substrate, a small volume of liquid is displaced and forms an initial meniscus of strong negative curvature. Because of its low capillary pressure, it then sucks the liquid from the film, rises on the fiber, and grows \cite{lestime2021dynamique,clanet_quere_2002,guo2019onset}.
This suction creates a groove corresponding to the apparition of a minimum in the thickness profile $h(r,t)$ is characterized by its thickness $h_\text{g}(t)$. Such a groove has been observed in many different situations and geometries as soon as a flat film is in contact with a meniscus: in soap films and soap bubbles, where the groove is also referred to as a dimple \cite{platikanov1964experimental,Aradian2001,atasi2020,gros2021,miguet2021marginal}, in coalescence problems \cite{chan11}, as well as in inkjet printing \cite{hack2018printing} or in the case of solid objects on liquid deposited films \cite{garcia2023drawing,lestime2021dynamique}.

Aradian et al. \cite{Aradian2001} predicted the self-similar profiles reached by the groove at long time in a soap film or a wetting film, in the vicinity of a meniscus of constant radius. 
In the situation of a vertical fiber, the meniscus radius grows over time and its negative Laplace pressure decreases in absolute value. Aradian's model therefore does not apply in this case, but the physical processes leading to a groove still hold. 
As the equilibrium profile reached at long time does not exhibit any minimum, the groove is only present in the transient regime. Consequently, $h_{\rm{g}}(t)$ has a non-monotonic time dependency, reaching its minimum value $h_{\rm{g}}^{\rm{min}}$ at $t_{\rm{min}}$.  
In this scenario, after $t_{\rm{min}}$,
the film thinning stops, and the perturbation induced by the dust particle relaxes towards a meniscus of height comparable to the dust particle size.
If $h_{\rm{g}}^{\rm{min}}$ is smaller than $h_{\rm{vdW}}=100~\rm{nm}$ at which intermolecular forces start to play a role.  
The precise value of $h_{\rm{g}}^{\rm{min}}$, and its comparison with $h_{\rm{vdW}}$, thus determines the fate of the final coating.

\begin{figure}[!ht]
    \centering
    \includegraphics[width=8.8cm]{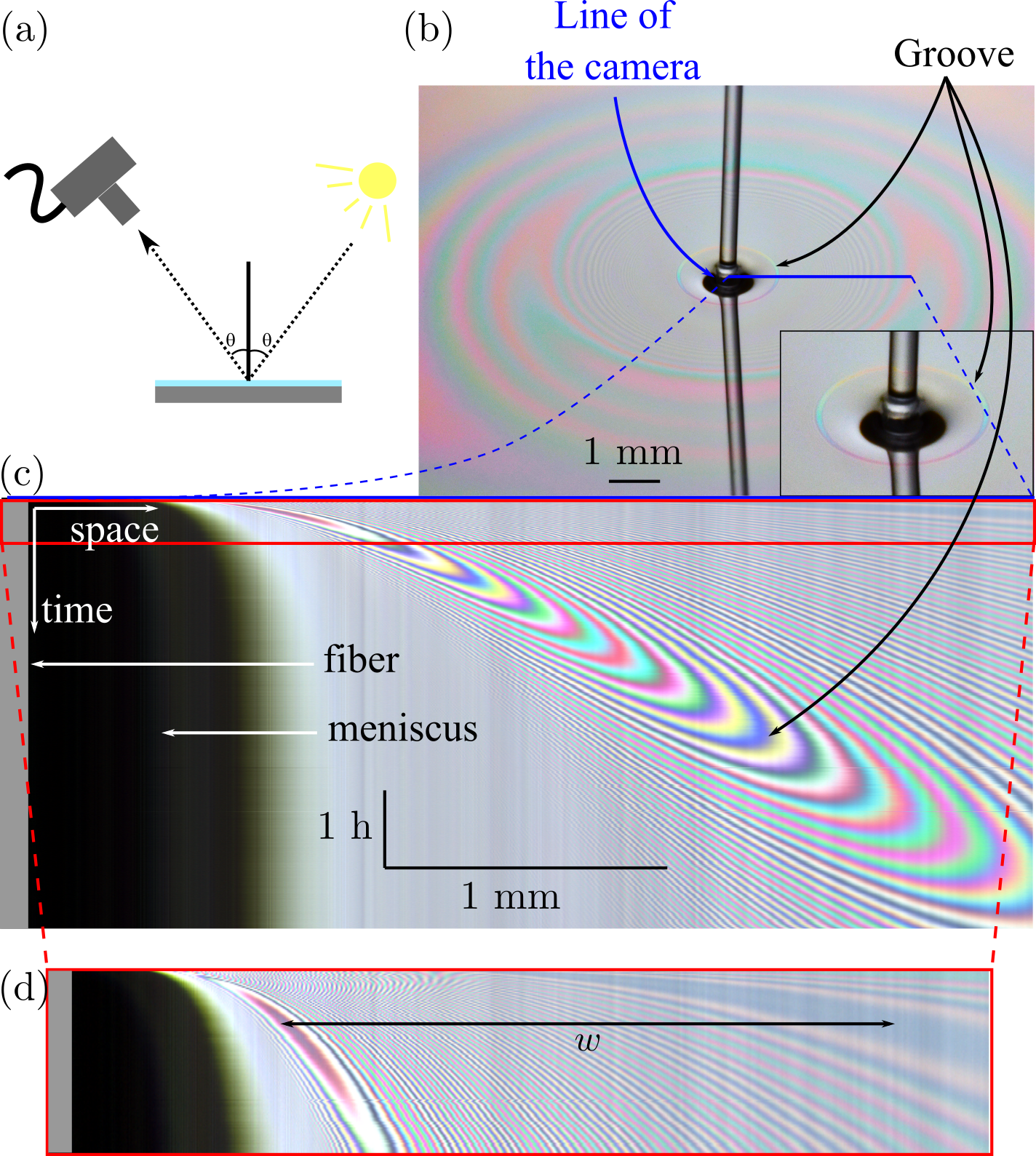}
    \caption{
    (a) Experimental setup: a hyperspectral camera receives the reflection of a white LED light on a thin silicone oil film deposited on a wafer.
    (b) Photograph of the film after depositing the fiber on the wafer. The lower half of the fiber is a reflection on the wafer. One can see the surface deformation thanks to the thin film interferences and especially the colored circle that detaches from the rest, which corresponds to the groove. The observation line of the camera, materialized by a blue solid line, is set to be a radius of the system.
    (c) Space-time diagram of the colors of the film:
    the radius of the fiber is drawn in grey, the black region is the meniscus, next to it there is a grey region, where the thickness cannot be measured because the spectra are not resolved due to high thickness variation in space, then there is a thin colored region which is the groove, and then a series of fringes which correspond to the rest of the film.
    (d) Crop at early time of the space-time diagram above. The first colored region, the closest to the meniscus, where the colors do not vary much, is the groove. The pink region corresponds to the time minimum of the groove thickness, $h_{\rm{g}}^{\rm{min}}$. The second flat colored region is the bump. The distance $w$ between the bottom of the groove and the top of the bump is highlighted.}
    \label{figure1}
\end{figure}

In this work, we provide the first experimental evidence of the time-minimum groove thickness.
We develop scaling laws for $h_{\rm{g}}^{\rm{min}}$ and $t_{\rm{min}}$ in good agreement with our quantitative measurements.
Additionally, we predict the critical film thickness below which the film will reach intermolecular repulsion forces.

\section{Experimental setup and results}

To observe the surface deformation in the vicinity of a cylindrical defect, we use a thin silicone oil film (HMS 301 Gelest) of viscosity $\eta=30~\rm{mPa.s}$ (at $25^\circ \rm{C}$), surface tension $\gamma=20~\rm{mN/m}$ and density $\rho = 980~\rm{kg/m}^3$. The capillary length is thus $l_{\rm{c}}=\sqrt{\gamma/(\rho g)} =1.4$~\rm{mm}. A film of controlled thickness is deposited on a silicon wafer by spin-coating (Polos Spin 150i) at a rotation speed between 400 and 780 rpm during 30 seconds. 
We explore a range of initial thicknesses $h_0$ from $9~\rm{\mu m}$ to $21~\rm{\mu m}$. 

A vertical cylindrical glass fiber of radius $r_{\rm{f}}$ (Hilgenberg, radii from $100~\mu \rm{m}$ to $500~\mu \rm{m}$) is fixed above the wafer, which is moved up with a lab jack.
The wafer is covered by a petri dish lid, in which a hole is drilled for the fiber to go through. 
The slow approach is stopped when the fiber begins to buckle, which makes sure that it is in contact with the wafer.
As soon as this contact is made, the interference patterns provide a direct signature of the surface deformation (Fig. \ref{figure1}(b)). The circle of bright colors indicated by an arrow in Figure \ref{figure1}(b) corresponds to the groove. 
This direct visualization shows that this groove remains axisymmetric, consistently with the good reproducibility of our groove measurements close to $t_{\rm{min}}$.
Nevertheless, for the thicker fibers, the buckling can result in a long-range asymmetric perturbation.
Moreover, spin-coating results in a thicker film at the edge of the wafer, which will cause heterogeneities far from the fiber.

\begin{figure}[t!]
    \centering
    \includegraphics[width=8.8cm]{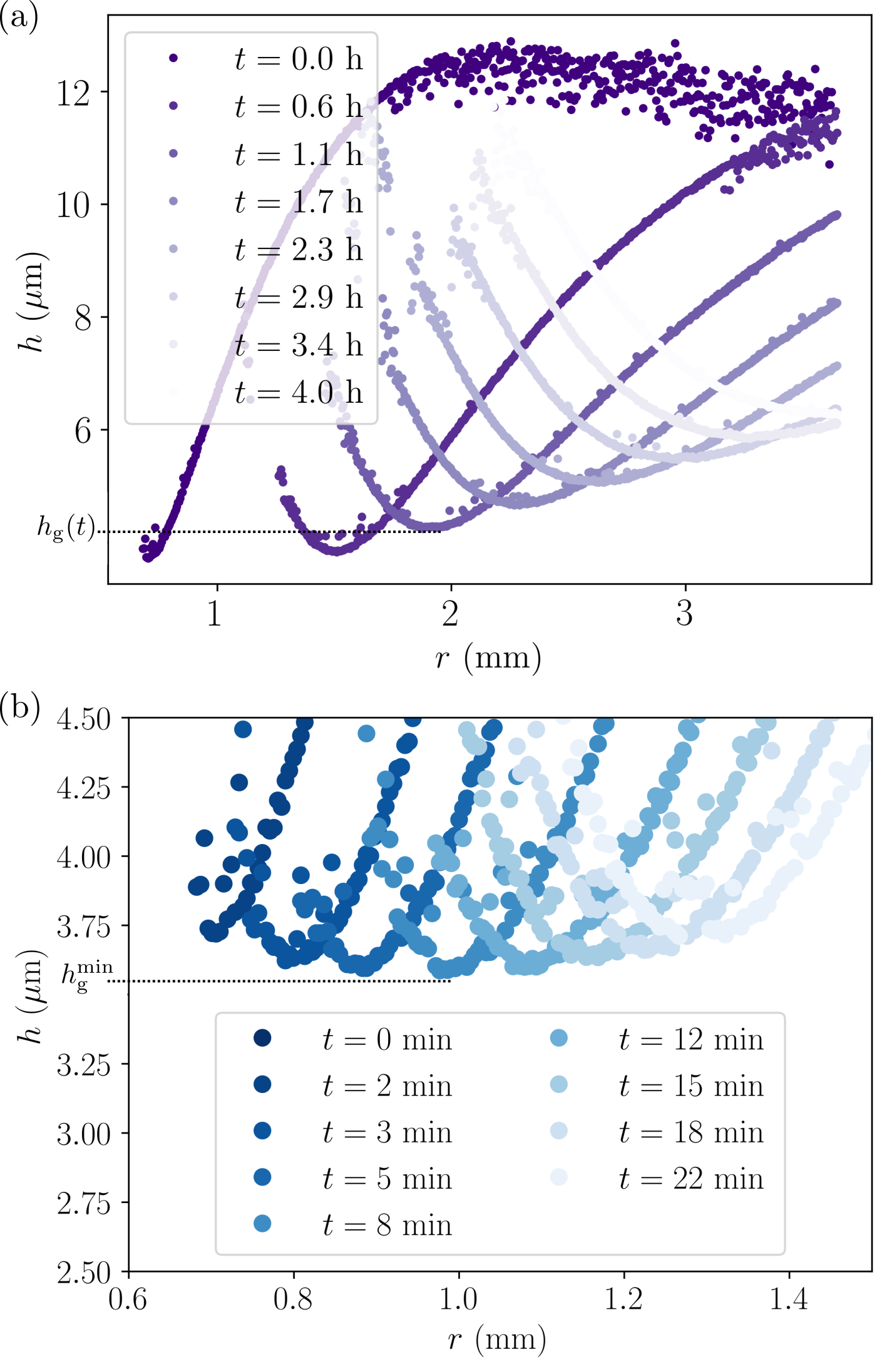}
    \caption{(a) Typical thickness profiles: thickness of the film $h$ as a function of the radial position $r$ for different times (here for $h_0=14.3~\rm{\mu m}$ and $r_{\rm{f}}=100~\rm{\mu m}$).
    $t$ is known to within 5 s, given the framerate of the camera. Close to the fiber, the thickness decreases as we go down the meniscus. The film gets thinner till it reaches a minimum $h_{\rm{g}}(t)$ at the bottom of the groove, and finally relaxes towards the flat part of the film, of thickness $h_0$. The thickness is not measured at small $r$, which corresponds to the meniscus, because the profile is too steep.
    (b) Thickness profiles at early times zoomed in on the groove: the groove first deepens and then fills back up.
    }
    \label{figure2}
\end{figure}

We measure quantitatively the film thickness along a line (blue solid line in Figure \ref{figure1}(b)) thanks to a hyperspectral camera (Pika L - Resonon), receiving the reflection of a white LED fiber optic light (Dolan-Jenner Fiber-Lite Mi-LED) on the film (Figure \ref{figure1}(a)). The camera is carefully oriented so that the line of measurement is along a radial direction.
A hyperspectral camera is the equivalent of a line of spectrometers, which measures the spectrum of light for each point of the line the camera is aiming at. 
The camera software produces a space-time diagram of the colors of the film (Figure \ref{figure1}(c)), with a light spectrum available at each pixel. 
We use the \textit{oospectro} python library \cite{oospectro} to process the obtained spectra and compute the corresponding thicknesses with a precision of $\pm 20$ nm.
The range of initial thicknesses that we can explore is limited. 
The camera spectral resolution is too low to resolve the numerous oscillations of the spectra at higher thicknesses. The \textit{oospectro} method prevents the measure of too small thicknesses , under 400 nm approximately.
The possibility to measure the entire profile at all times determines our choice of initial film thickness range: between $h_0=8.4~\mu \rm{m}$ and $h_0=21 ~\mu\rm{m}$.
We obtain the profile $h(r)$,  with $r$ the radial distance from the center of the fiber, determined as the middle of the black region at $t=0$. $r$ is known with a precision of $5~\mu\rm{m}$ which is the resolution of the camera. In Figure \ref{figure1}(c) the space-time diagram is cropped so that the first black pixel on the left corresponds to the edge of the fiber.

\begin{figure}[t!]
    \centering
    \includegraphics[width=8.8cm]{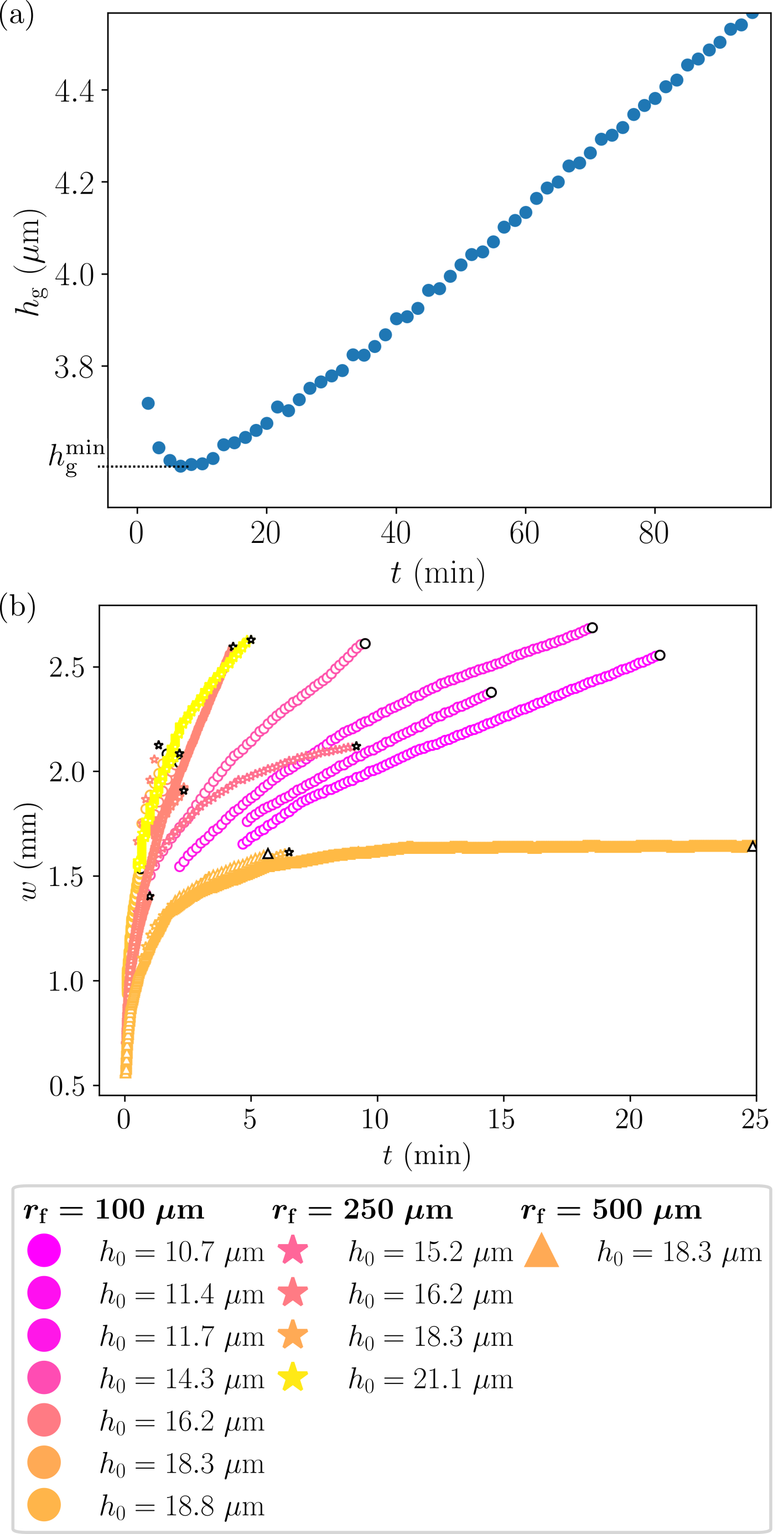}
    \caption{(a) Typical time evolution of the thickness of the groove $h_{\rm{g}}$ (here for $h_0=14.3~\rm{\mu m}$ and $r_{\rm{f}}=100~\rm{\mu m}$). First, the groove deepens, then it reaches a minimum, and finally, it fills back up again.
    (b) Width of the groove $w$ over time for different initial thicknesses $h_0$ and fiber radii $r_{\rm{f}}$: the groove gets wider with time and with thicker films. The black symbols are meant to distinguish better the shapes of the symbols.
    }
    \label{figure3}
\end{figure}

In Figure \ref{figure2}(a), we plotted different thickness profiles for different times ranging from $t=0$ h to $t=4.0$ h.
The origin of time is determined when the perturbation appears on the space-time diagram, the acquisition starting before making the contact.
The precision of $\pm 5~\rm{s}$ is given by the framerate of the camera. For all the times, we observed similar shapes for the profiles.
At small $r$, \textit{i.e.} close to the fiber (all the geometrical notations are defined in Figure \ref{figure4}), and in the upper part of the meniscus, there are no points as the meniscus is too steep to measure the thickness.
At the end of the meniscus, the thickness $h(r)$ decreases, goes through a minimum, and increases again. 
This space minimum corresponds to the bottom of the groove at the position $r_{\rm{g}}$ and of thickness $h_{\rm{g}}(t)$.
On the first recordable profile shown in Figure \ref{figure2}(a), at $t = 0~h$, we can see that at $r$ larger than $r_{\rm{g}}$, the thickness increases and reaches a maximum (a bump) at $r_{\rm{b}}$ and then relaxes towards the flat part of the film. 
The positions $r_{\rm{g}}(t)$ and $r_{\rm{b}}(t)$ increase with time. 
Focusing on the groove at early times, in Figure \ref{figure2}(b), we see that the groove first deepens until a thickness $h_{\rm{g}}^{\rm{min}}$ and then fills back up. 
To quantify this space minimum, we extract the thickness at the bottom of the groove $h_{\rm{g}}(t)$ from those thickness profiles and plot it against time in Figure \ref{figure3}(a).
We see that the evolution of the groove thickness $h_{\rm{g}}(t)$  is indeed non-monotonic and reaches its time-minimum value $h_{\rm{g}}^{\rm{min}}$ at the time $t_{\rm{min}}$.
This is the time-minimum thickness we mentioned in the introduction. In the discussion we will focus on this time-minimum thickness of the groove, $h_{\rm{g}}^{\rm{min}}$ and show that it is governed by the growth of the meniscus.
In the following, all parameters $X$ taken at this instant will be noted $X_{\rm{min}}$ or $X^{\rm{min}}$.

We can also extract the "width" of the groove $w$, defined as the distance between the bottom of the groove and the top of the bump, $r_{\rm{b}} - r_{\rm{g}}$.
The measured value is extracted from the space-time diagrams as shown in Fig. \ref{figure1}(d).
However, we can only measure $w$ at early times ($\sim 10~\rm{min}$) given that $w$ grows over time (Figure \ref{figure3}(b)) and the bump quickly goes out of the field of view of the camera.

\section{Discussion} 

\begin{figure}
    \centering
    \includegraphics[width=8.8cm]{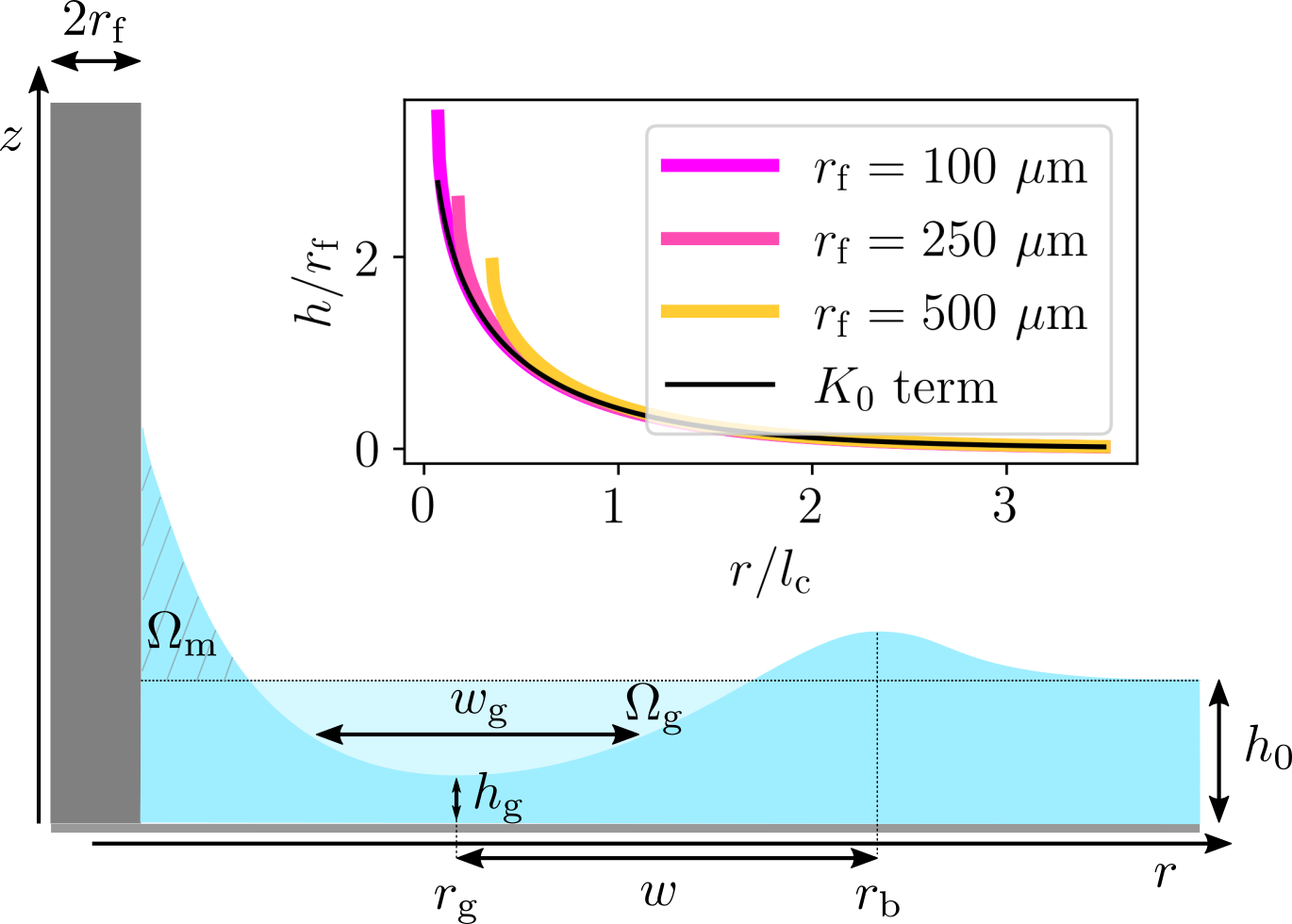}
    \caption{Notations used in the models: $r_{\rm{f}}$ is the fiber radius, $h_0$ is the initial thickness of the film, $w$ is the width of the groove, $r_{\rm{g}}$ is the position of the groove, $r_{\rm{b}}$ is the position of the bump, $\Omega_{\rm{m}}$ is the volume of the meniscus, $\Omega_{\rm{g}}$ is the volume of the groove.
    (inset): Profile of the static meniscus computed with the equation of the interface from James \cite{james_1974} (section 6). The black line is the asymptotic profile for $r \gg r_{\rm{f}}$.
    }
    \label{figure4}
\end{figure}

Let us start with a description of the equilibrium shape of the meniscus.
The height of the meniscus is fixed by the fiber radius $r_{\rm{f}}$.
Nevertheless, even for small fiber radii, the lateral extension far from the fiber is given by the capillary length $l_{\rm{c}}$ \cite{james_1974,clanet_quere_2002}.
The semi-analytic expression of the entire profile in the limit of fiber of radius much smaller than the capillary length is $h^{\rm{J}}(r)$, given in section 6 from reference \cite{james_1974}. 
\begin{equation}
h^{\rm{J}}(r)=r_{\rm{f}}\left[\mbox{ln}\left(\frac{2 r}{r+ \sqrt{r^2-r_{\rm{f}}^2}} \right) + K_0\left(\frac{r}{l_{\rm{c}}}\right) \right] \, ,
\label{james}
\end{equation}
where $K_0$ is the modified Bessel function of the second kind of order 0.
This profile is plotted in the inset of Fig. \ref{figure4} for the three fiber radii we use.
Its asymptotic behavior at a distance $r \gg r_{\rm{f}}$  is $h^{\rm{J}} \sim r_{\rm{f}} K_0(r/\ell_{\rm{c}})$.
We checked (not shown) that the equilibrium meniscus volume is dominated by this long-range behavior and, using $\int_0^\infty x K_0(x) \,dx = 1$, we get
\begin{equation}
    \Omega_{\rm{m}}^{\rm{eq}} \sim r_{\rm{f}} l_{\rm{c}}^2.
    \label{eq:VolumeMenisque}
\end{equation}
The scaling of the radius of curvature of the static meniscus $r_{\rm{rz, eq}}$ in the ($r$,$z$) plane will also be needed in the following. Close to the groove minimum, it is dominated by the last term of eq. \ref{james} and scales as :
\begin{equation}
  \frac{1}{r_{\rm{rz, eq}}} \sim \frac{r_{\rm{f}}}{l_{\rm{c}}^2}.
   \label{courb_eq}
\end{equation}

\begin{figure}[h!]
    \centering
    \includegraphics[width=8.8cm]{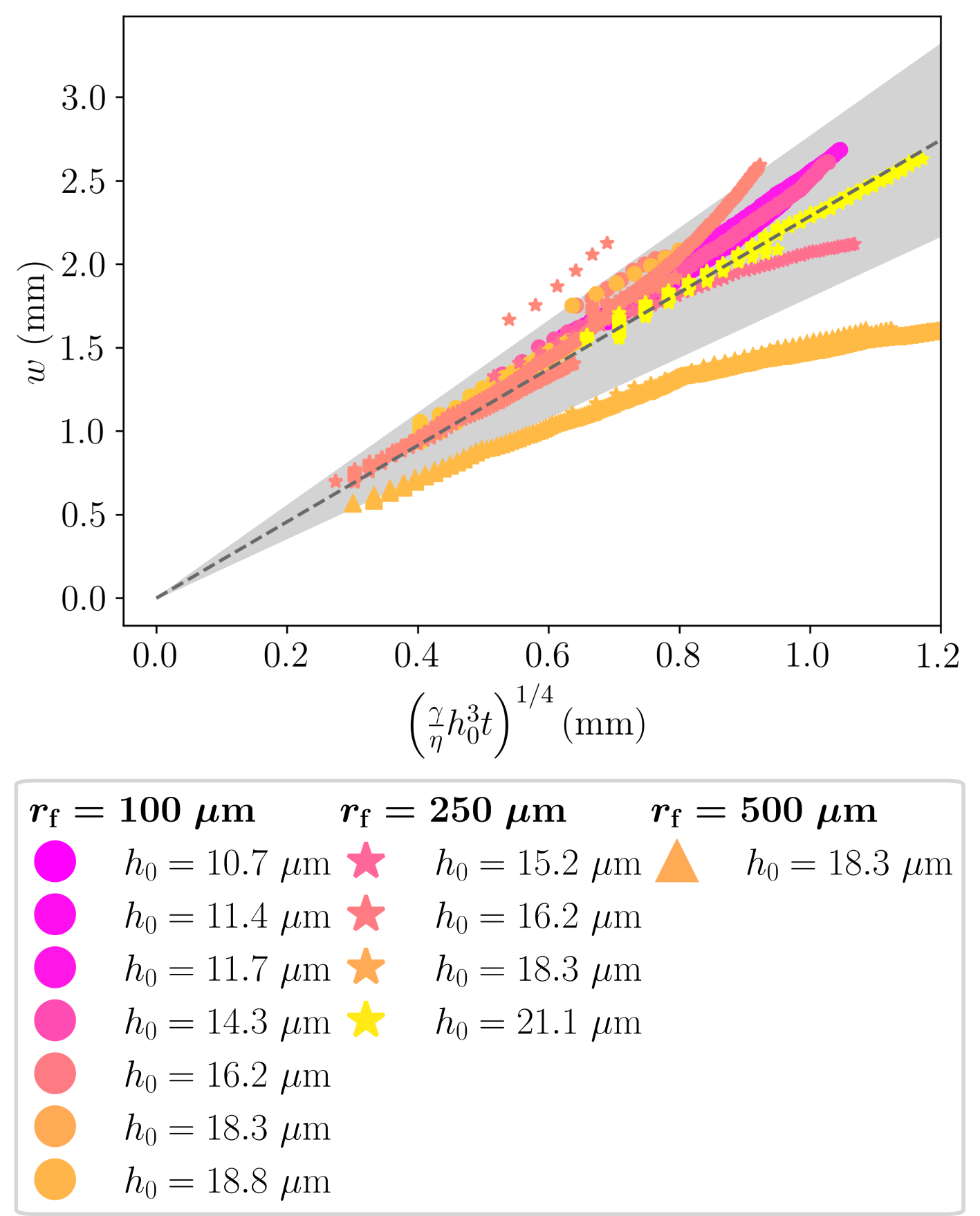}
    \caption{Width of the groove $w$ as a function of the scaling of $w$ for different fiber radii $r_{\rm{f}}$ and initial thicknesses $h_0$: the data collapse on a master curve with a bit of scattering due to experimental conditions. The dashed line is the mean of the slopes resulting from the fits of each experiment; the coefficient of this slope is $2.2$. The upper and lower slopes of the grey zone correspond to the slope of the dashed line plus or minus the standard deviation of all the slopes of the fits of each experiment, this standard deviation being 0.45.}
    \label{figure5}
\end{figure}

Before reaching its equilibrium shape, the thin film is governed by the lubrication equation
\begin{equation}
    \frac{\partial h}{\partial t} =  \frac{\gamma}{3\eta} \nabla \cdot \left(h^3 \nabla\nabla^2h\right).
    \label{ThinFilmEquation} 
\end{equation}
With a vertical characteristic length scale $h_0$, this equation directly imposes a scaling for the in-plane length scale, which has been proven to be valid in many different systems \cite{hack2018printing,lakshman2021deformation,garcia2023drawing,MacGraw2012}.
This scaling
\begin{equation}
    w\sim\left(\frac{\gamma}{\eta} h_0^3 t\right)^{1/4},
    \label{eq:Scaling_W}    
\end{equation}
is also observed to rescale our data (Figure \ref{figure5}).
The different curves roughly collapse on a single master curve of slope $2.2 \pm 0.45$.
For the fiber of radius $250~\rm{\mu m}$, the data are more dispersed than for the fiber of radius $100~\rm{\mu m}$. 
This is due to the fact that for several experiments for $r_{\rm{f}}=250~\rm{\mu m}$, the perturbation is not perfectly axisymmetric around the fiber as previously discussed. 
Moreover, the edge effects discussed previously are probably the cause of some curves departing from the straight line at late time. 

However, in our specific situation, the film is in contact with a meniscus, whose radius of curvature $r_{\rm{rz}}$ introduces another length scale. The above scaling (Eq. \ref{eq:Scaling_W}) is therefore not the only possible one for in-plane lengths, and a more detailed study is required. 
The long-time behavior of an initially flat film in contact with a meniscus has been solved by Aradian \textit{et al.} \cite{Aradian2001}. In their case, the meniscus is invariant by translation in one direction and of constant radius of curvature. The scalings of our data will be obtained by adapting their results to our specific geometry.

In the geometry of Aradian \textit{et al.}, the groove can be described through three characteristic length scales (denoted respectively $l$, $h$ and $w$ in \cite{Aradian2001} and adapted to our notations in the following, with a superscript 'a'): 
the lateral extension of the perturbation $w^{\rm{a}}$, the thickness at the minimum of the groove $h_{\rm{g}}^{\rm{a}}$ and the in-plane characteristic length near the bottom of groove $w_{\rm{g}}^{\rm{a}}$ (typically the groove width at $h=2 h_{\rm{g}}^{\rm{a}}$).
The scaling obtained for $w^{\rm{a}}$ does not depend on the presence of the meniscus and the scaling of Eq. \ref{eq:Scaling_W} is recovered, consistently with our experimental observations.
However, $w_{\rm{g}}^{\rm{a}}$ scales differently and is obtained by (i) matching the film curvature on the left side of the groove with the meniscus curvature and (ii) matching the slopes on the right side of the groove. 
As the meniscus size depends on the time in our situation, the time scalings for $w_{\rm{g}}^{\rm{a}}$ and $h_{\rm{g}}^{\rm{a}}$ cannot be directly used.
In particular, with a steady meniscus, the groove thickness $h_{\rm{g}}^a$ is a decreasing function of time whereas in our situation $h_{\rm{g}}(t)$ is non-monotonic. 
However, the matching conditions (i) and (ii) are true at each time and should remain valid in our case, leading to, respectively, 
\begin{equation}
    \frac{h_{\rm{g}}}{{w_{\rm{g}}}^2} \sim \frac{1}{r_{\rm{rz}}},
    \label{eq:MatchingCurvature}
\end{equation}
and 
\begin{equation}
    \frac{h_{\rm{g}}}{w_{\rm{g}}} \sim \frac{h_0}{w}.
    \label{eq:MatchingSlopes}
\end{equation}
These scalings impose that, at each time, the minimal thickness in the groove scales as
\begin{equation}
     h_{\rm{g}} \sim r_{\rm{rz}} \left(\frac{h_0}{w}\right)^2
    \label{hg_theo}
\end{equation}

\section{Scaling laws for $h_{\rm{g}}^{\rm{min}}$ and $t_{\rm{min}}$}

We propose to associate the conditions of Eq. \ref{hg_theo}  with a volume conservation law to predict $h_{\rm{g}}^{\rm{min}}$.
At time $t_{\rm{min}}$, the capillary suction of the meniscus is overcome by the leveling process of the film, and we assume that 
the meniscus shape already scales as the equilibrium shape, so, using Eq. \ref{eq:VolumeMenisque},  $\Omega_{\rm{m}}^{\rm{min}} \sim r_{\rm{f}} l_{\rm{c}}^2 $.
Indeed, in Figure \ref{figure1}(c), one can see that at $t_{\rm{min}}$ the black zone, which is the meniscus width, has already reached 75\% of its equilibrium value.
On the other hand, the volume extracted from the groove at this time can be estimated as
\begin{equation}
    \Omega_{\rm{g}}^{\rm{min}} \sim \pi \left(({r_{\rm{b}}^{\rm{min}}})^2 - ({r_{\rm{g}}^{\rm{min}}})^2\right) (h_0 - h_{\rm{g}}^{\rm{min}}),
\end{equation}
with $r_{\rm{g}}^{\rm{min}}$ and $r_{\rm{b}}^{\rm{min}}$ respectively the positions of the groove and the bump at $t_{\rm{min}}$ (Figure \ref{figure4}).
We assume that the groove is very deep so $h_0 - h_{\rm{g}}^{\rm{min}} \sim h_0$ and that $w_{\rm{min}} \sim r_{\rm{b}}^{\rm{min}}$, which leads to:
\begin{equation}
    \Omega_{\rm{g}}^{\rm{min}} \sim h_0 w_{\rm{min}}^2.
\end{equation}
The volume conservation imposes that this volume extracted from the groove equals the sum of the meniscus and bump volumes. 
Assuming that the bump volume remains smaller than the meniscus volume (or at most of the same order), we obtain that 
$\Omega_{\rm{m}}^{\rm{min}}$ scales as $\Omega_{\rm{g}}^{\rm{min}}$.
The volume conservation thus writes $r_{\rm{f}} l_{\rm{c}}^2 \sim h_0 w_{\rm{min}}^2$ which gives
\begin{equation}
    w_{\rm{min}} \sim l_{\rm{c}}\left(\frac{r_{\rm{f}}}{h_0}\right)^{1/2}.
    \label{eq:Scaling_Wmin}
\end{equation}
This cannot be tested directly as the bump is outside the field of view of the camera at $t_{\rm{min}}$, so we cannot measure $w$ any longer.

\begin{figure}[t!]
    \centering
    \includegraphics[width=8.8cm]{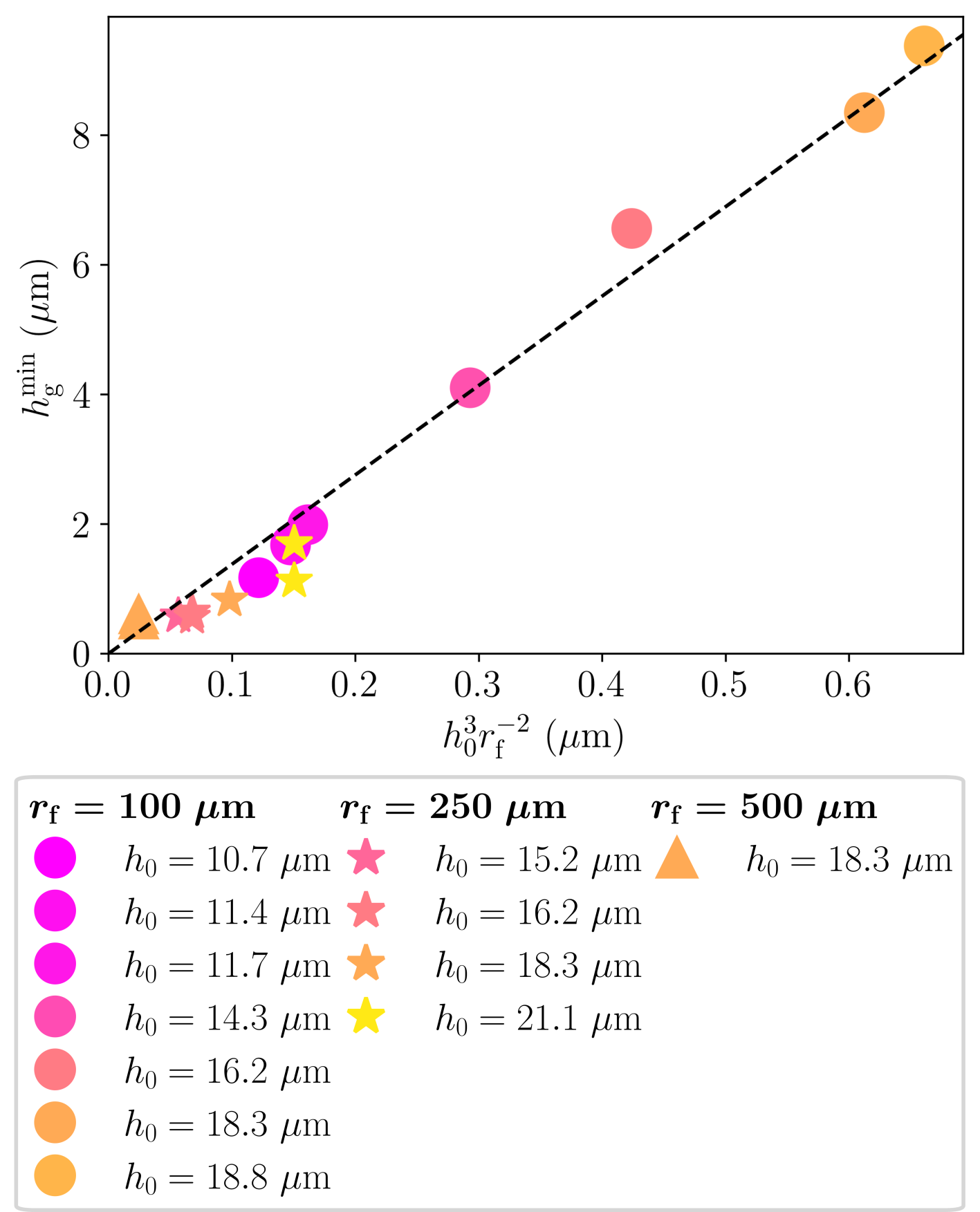}
    \caption{Time minimum of the groove thickness $h_{\rm{g}}^{\rm{min}}$ as a function of the theoretical value of $h_{\rm{g}}^{\rm{min}}$ for different fiber radii rf and initial thicknesses h0. The dashed line is a fit of all the data, the coefficient is 14.
    }
    \label{figure7}
\end{figure}

We use Eq. \ref{hg_theo} at $t_{\rm{min}}$, Eq. \ref{eq:Scaling_Wmin}, and the scaling of $r_{\rm{rz}}$ at equilibrium  (Eq. \ref{courb_eq}) 
to build a scaling law for $h_{\rm{g}}^{\rm{min}}$: 
\begin{equation}
    h_{\rm{g}}^{\rm{min}} \sim \frac{h_0^2 l_{\rm{c}}^2}{w_{\rm{min}}^2 r_{\rm{f}}} \sim\frac{h_0^3}{r_{\rm{f}}^2}.
    \label{eq:hgmin}
\end{equation}
This scaling law describes well our data of the time-minimum thickness of the groove for different initial thicknesses and fiber radii (Figure \ref{figure7}) as they collapse on a master curve.
The dashed line is a fit of all the data and its slope is 2 which is of order unity, therefore the scaling law is validated.
\begin{figure}[t!]
    \centering
    \includegraphics[width=8.8cm]{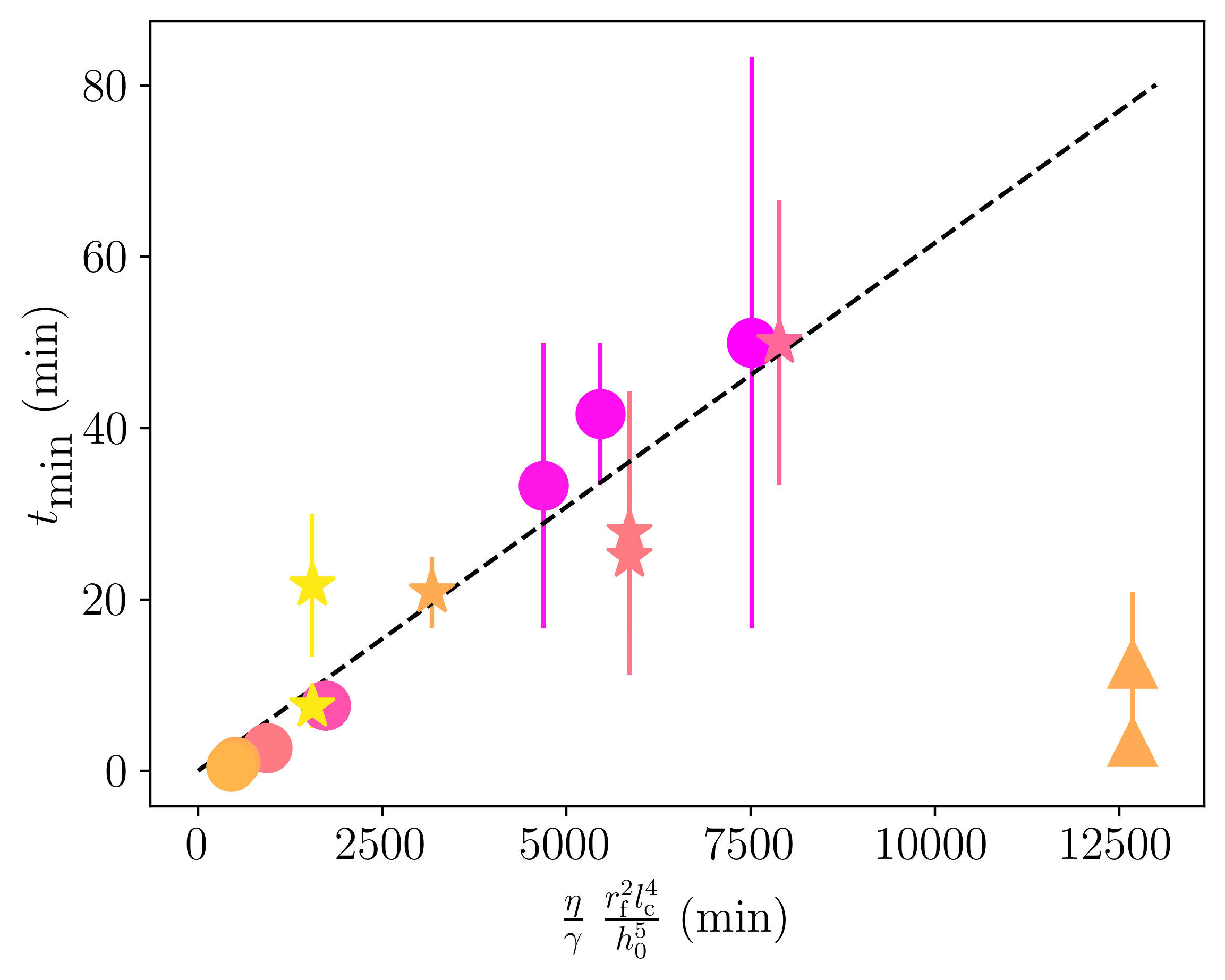}
    \caption{
    Time $t_{\rm{min}}$ when the groove reaches its minimum thickness, as a function of the theoretical value of $t_{\rm{min}}$, for different fiber radii $r_{\rm{f}}$ and initial thicknesses $h_0$.
    All data collapse on a master curve except the data for the largest fiber radius $r_{\rm{f}}=500~\mu\rm{m}$ as this radius gets close to the capillary length therefore the length scale separation begins to fail. 
    The dashed line is a fit of all the data except $r_{\rm{f}}=500~\rm{\mu m}$, the coefficient being $1/162$.
    The error bars are estimated based on the width of the time interval on which we determine the time-minimum thickness.
    The legend is the same as in Figure 6 on the left.
    }
    \label{figure6}
\end{figure}

Additionally, assuming that the scaling of $w$, given by  Eq. \ref{eq:Scaling_W},  still holds at $t_{\rm{min}}$, and combining it  with Eq. \ref{eq:Scaling_Wmin}, we get a scaling for $t_{\rm{min}}$:
\begin{equation}
    t_{\rm{min}} \sim \frac{\eta}{\gamma}\frac{r_{\rm{f}}^2 l_{\rm{c}}^4}{h_0^5}.
    \label{eq:tmin}
\end{equation}

This scaling is tested in Figure \ref{figure6}.
The data for different initial thicknesses $h_0$ and fiber radii $r_{\rm{f}}$ collapse on a master curve for the smallest fiber radii $r_{\rm{f}} = 100 \, \mu$m and 250 $\mu$m.
The dashed line is a straight line fit of these two data series.
The surprisingly small slope of $1/162$ could be attributed to the rough estimation of the volumes.

The data obtained with the fiber of radius $r_{\rm{f}} = 500 \,  \mu$m are significantly below the master curve. In that case, $r_{\rm{f}}$ gets close to the capillary length ($l_{\rm{c}}=1.4~\rm{mm}$), and the length scale separation begins to fail, which may explain this discrepancy.

\section{Towards intermolecular repulsion}

\begin{figure}[t!]
    \centering
    \includegraphics[width=8.8cm]{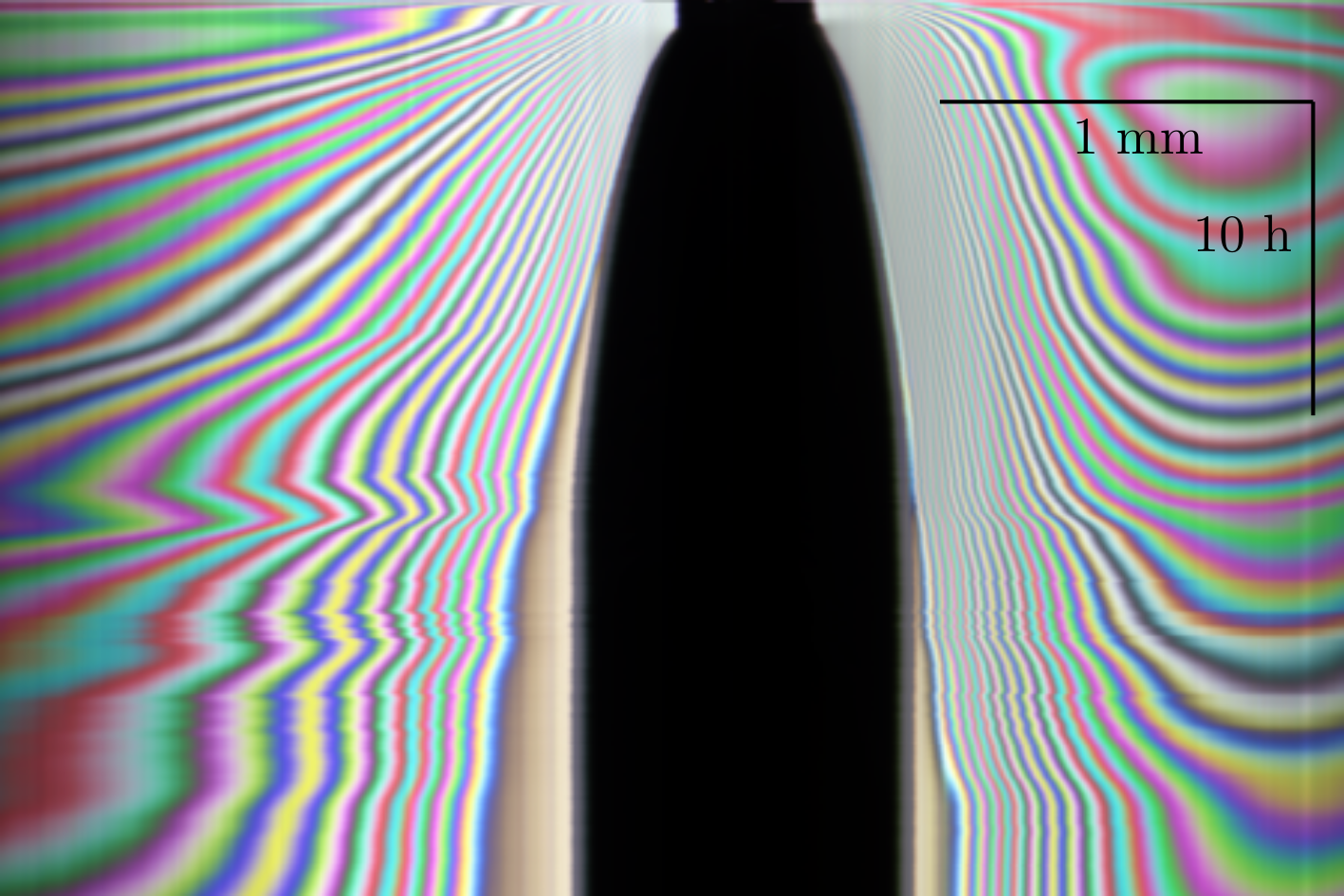}
    \caption{Space-time diagram of a film which exhibits a Newton film before $t_{\rm{min}}$.
    The radius of the fiber is $250~\rm{\mu m}$ and the initial thickness of the film is $5~\rm{\mu m}$.
    Close to the meniscus, we see a beige fringe that gets much larger than the other fringes.
    Before the beige fringe enlarges, there is a groove very close to the meniscus, which is difficult to see on the image.
    }
    \label{fig:stdem}
\end{figure}

As $h_{\rm{g}}^{\rm{min}}$ decreases if $h_0$ decreases, if we make experiments with even smaller initial thicknesses (smaller than $10~\rm{\mu m}$), we expect $h_{\rm{g}}^{\rm{min}}$ to reach a thickness at which the intermolecular forces start to play a role before or at $t_{\text{min}}$.
To reach such small thicknesses, we have used a silicon oil with a smaller viscosity, $\eta = 3$ mPa.s.
Indeed, the process is so slow at small thicknesses with the viscous oil that the heterogeneities due to the spin-coating start to invade the film before $t_{\text{min}}$.
For a less viscous oil, we sometimes obtain space-time diagrams that look like the one in Figure \ref{fig:stdem}.
At early time, there is a groove very close to the meniscus; its colors are barely visible on the space-time diagram.
At longer time, we see that a zone of almost uniform color (beige here) appears close to the meniscus which is the black area. 
It corresponds to a thin flat zone.
To measure the thickness in this zone, we use the spectra measured by the hyperspectral camera. The thickness is too small to use the \textit{oospectro} library to compute the thickness. Thus, we use the Sheludko renormalization \cite{Sheludko}. It consists of inverting the function giving the intensity as a function of the thickness since it is bijective for small enough thicknesses. The measured thickness is of the order of 10 nm.
We call this thin film a Newton film in reference to the Newton black film in soap films, which necessitate repulsive forces to be stable.

\begin{figure}[t!]
    \centering
    \includegraphics[width=8.8cm]{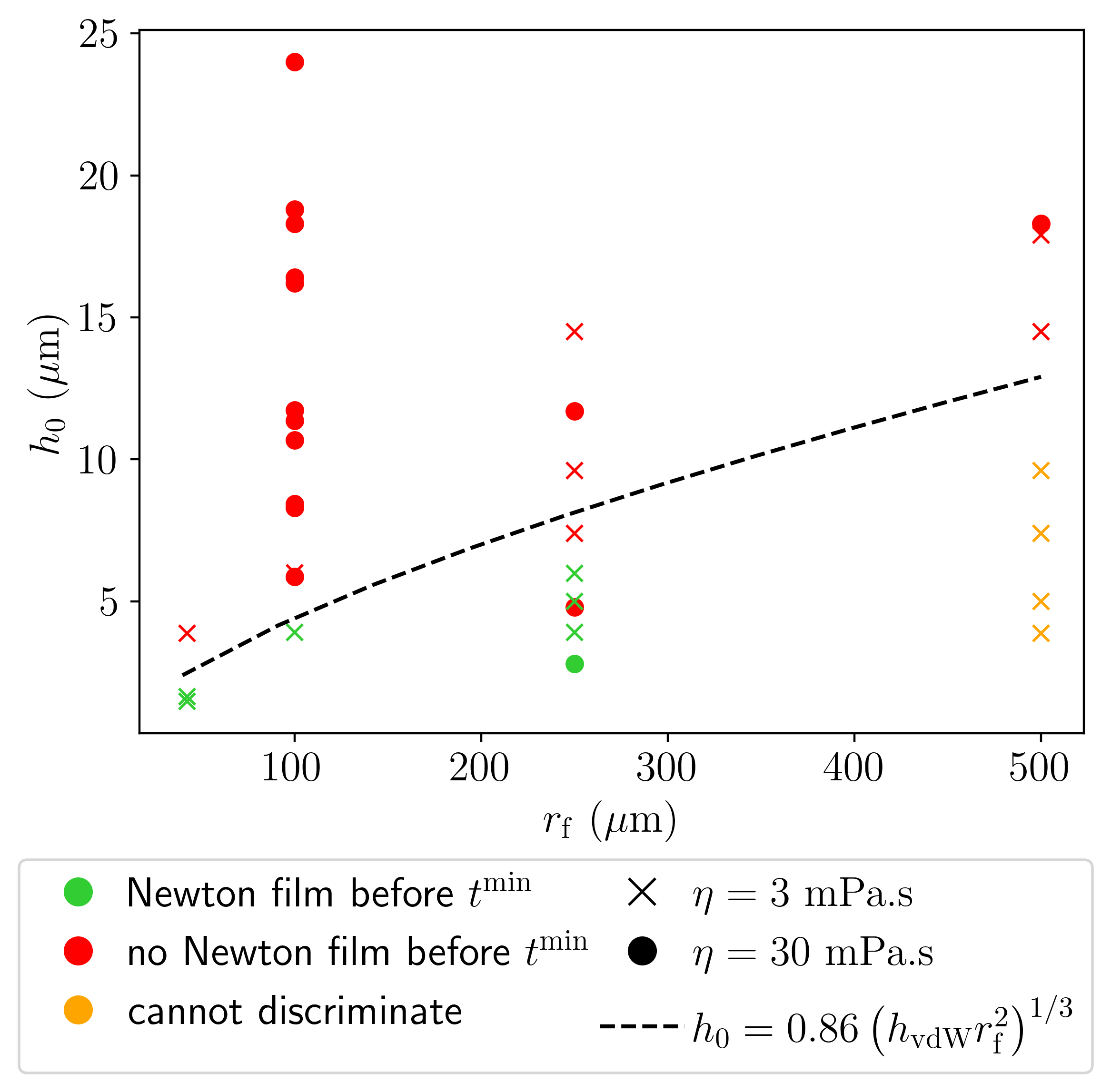}
    \caption{Phase diagram of the film: the initial thicknesses $h_0$ are plotted against the fiber radii $r_{\rm{f}}$. One point corresponds to one experiment. Crosses denote a viscosity $\eta = 3~\rm{mPa.s}$ (PDMS) while full points stand for $\eta = 30~\rm{mPa.s}$ (HMS 301 Gelest).
The green points correspond to situations in which the Newton film appears before the groove reaches its minimum thickness (at $t_{\rm{min}}$) and red points correspond to when there is no Newton film before $t_{\rm{min}}$.
    $t_{\rm{min}}$ is determined \textit{via} the scaling of Eq. \ref{eq:tmin} and the coefficient from Fig. \ref{figure6}.
    Orange points are for experiments with $r_{\rm{f}}=500~\rm{\mu m}$ where $t_{\rm{min}}$ could not be determined theoretically as the scaling law was not validated for this fiber radius.
    The $42.5~\rm{\mu m}$ radius fiber has been made by melting a $100~\rm{\mu m}$ radius fiber from Hilgenberg.
    The dashed line is what we expect for the scaling of Eq. \ref{eq:h0*}.
    The coefficient comes from the fit of Figure \ref{demouillagescaling}.
    }
    \label{demouillagediag}
\end{figure}

We want to measure the critical initial thickness for which the groove will deepen enough (before $t_{\rm{min}}$) to reach a thickness at which a Newton film appears.
We built a phase diagram, plotted in Figure \ref{demouillagediag}, discriminating between experiments where a Newton film appears before $t_{\rm{min}}$ (in green) or does not appear before $t_{\rm{min}}$ (in red).
$t_{\rm{min}}$ is determined \textit{via} the scaling of Eq. \ref{eq:tmin} and the coefficient from Fig. \ref{figure6}.
Orange points are for experiments with $r_{\rm{f}}=500~\rm{\mu m}$ where $t_{\rm{min}}$ could not be determined theoretically as the scaling law was not validated for this fiber radius.
We call $h_0^*$ the thickness at which we observe a transition between the red and the green phase.

\begin{figure}[t]
    \centering
    \includegraphics[width=8.8cm]{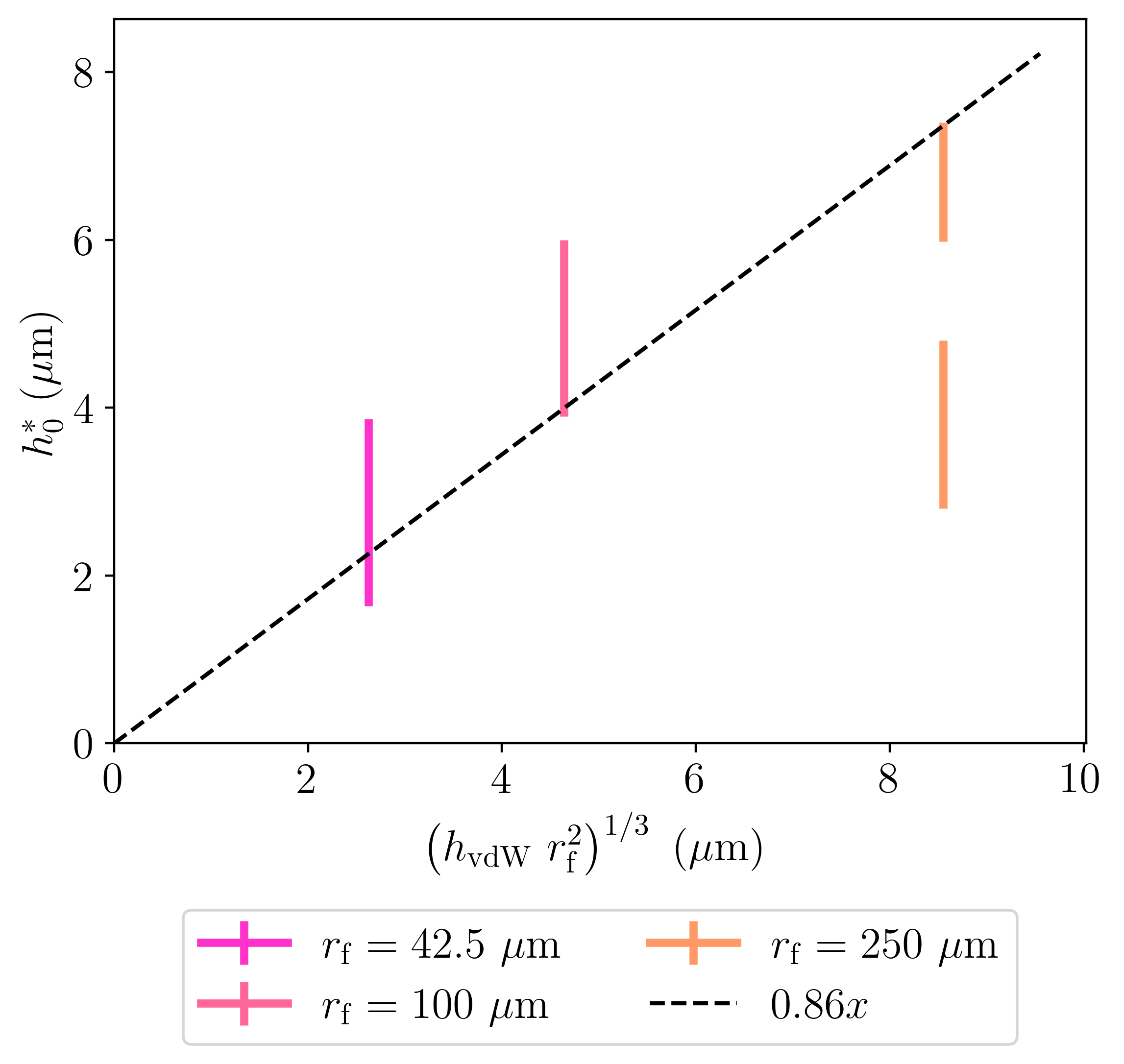}
    \caption{Critical initial thickness $h_0^*$ for reaching the Newton film as a function of its scaling law, for different fiber radii $r_{\rm{f}}$. The tops of the error bars are the minimum initial thicknesses $h_0$ for which the film exhibits a groove, and the bottoms of the error bars are the maximum initial thicknesses $h_0$ for which there is a Newton film. $h_{\rm{vdW}}$ is taken equal to $10~\rm{nm}$. The dashed line is a straight line fit of the middle of the error bars, which slope is $0.86$. The two lines for $r_{\rm{f}}=250~\rm{\mu m}$ come from the two different viscosities used.
    }
    \label{demouillagescaling}
\end{figure}

For silicone oil on a silicon wafer, the van der Waals forces between both interfaces are repulsive \cite{leger_liquid_1992,cazabat_dynamics_1997}, which may prevent the groove from thinning further.
The thickness thus saturates to $h_{\rm{vdW}}$, which explains the almost uniformly beige zone on the space-time diagram (Fig. \ref{fig:stdem}).
This situation is reached as soon as $h_{\rm{g}}^{\rm{min}} = h_{\rm{vdW}}$.
Using the scaling for $h_{\rm{g}}^{\rm{min}}$ (Eq. \ref{eq:hgmin}) gives the critical initial thickness:
\begin{equation}
    h_0^* \sim (h_{\rm{vdW}} r_{\rm{f}}^2)^{1/3}.
    \label{eq:h0*}
\end{equation}
This scaling law is tested in Figure \ref{demouillagescaling}. The data align on a straight line of slope $0.86$ for $h_{\rm{vdW}}=10~\rm{nm}$ (measured previously), which confirms the scaling.
This scaling law is also plotted in Figure \ref{demouillagediag} as a dotted line and is in reasonable agreement with the experimental data.

\section{Conclusion}
To summarize, we have measured the evolution of the thickness of the groove, which appears around the meniscus created by a vertical cylindrical fiber on a thin silicone oil film. 
This groove propagates radially over time and its thickness exhibits a minimum in time. 
The width of this groove $w$ is well described by a scaling law accounting for the flow which relaxes this perturbation using the thin film equation, as previously shown in various geometries. 
We build another scaling law for the time necessary to reach the time-minimum thickness of the groove assuming that the whole volume of the meniscus comes from the groove and that the time-minimum thickness of the groove is reached when the suction of the meniscus stops, that is to say at the equilibrium of the meniscus.
This scaling law is in very good agreement with the measured values of $t_{\rm{min}}$, except for the largest fiber because its radius gets comparable to the capillary length. 
Using geometric arguments on the shape of the groove, based on an asymptotic matching on each side of the groove, we obtain a scaling law for the time-minimum thickness of the groove $h_{\rm{g}}^{\rm{min}}$, which rescales our data.

For small initial thicknesses $h_0$, the groove reaches a thickness ($\sim 10~\rm{nm}$) at which intermolecular forces start to have an effect and the space-time diagrams exhibit a large zone where the thickness is uniform corresponding to a thin flat zone.
Based on the scaling law for $h_{\rm{g}}^{\rm{min}}$, we deduce a scaling law for the critical initial thickness $h_0^*$ below which disjoining repulsion forces will play a role.
Experimental data are well described by this prediction.

As for perspectives, the dynamics of the groove while going down and then up are still open questions, as well as the Newton film dynamics: enlargement velocity, influence of the nature of the substrate etc.
Moreover, drying could be added to the system to mimic the industrial coating problem.

\section*{Open data}
Data for this article are available at https://doi.org/10.5281/zenodo.12801627.

\section*{Author Contributions}
I. C., A. E.-S., F.R. and E.R. designed the experimental study. A. E.-S. carried out the experiments. I. C., A. E.-S., F.R. and E.R. analyzed and interpreted the experimental results and wrote the manuscript.

\section*{Conflicts of interest}
There are no conflicts to declare.

\section*{Acknowledgements}
This research was funded by the Agence Nationale de la Recherche (ANR), grant ANR-20-CE30-0019 (DRAINFILM). A CC BY license is applied to the AAM arising from this submission, in accordance with the grant’s open access conditions. We thank Arnaud Antkowiak and Théophile Gaichies for fruitful discussions.

\bibliographystyle{unsrt}

\end{document}